\documentclass[MSNbibl,nameyear,dvips]{arxstspdf}
\usepackage{graphicx}

\volume{26}
\issue{3}
\pubyear{2011}
\firstpage{299}
\lastpage{316}
\doi{10.1214/11-STS352}

\begin{document}
\begin{frontmatter}

\title{Is Bayes Posterior just Quick~and~Dirty~Confidence?\thanksref{T1}}
\relateddoi{T1}{Discussed in \doi{10.1214/11-STS352A},
\doi{10.1214/11-STS352B}, \doi{10.1214/11-STS352C} and \doi{10.1214/11-STS352D}; rejoinder at \doi{10.1214/11-STS352REJ}.}

\begin{aug}
\author[a]{\fnms{D. A. S.} \snm{Fraser}\corref{}\ead[label=e1]{dfraser@utstat.toronto.edu}}
\runauthor{D. A. S. Fraser}

\affiliation{University of Toronto and University of Western Ontario}

\address[a]{D. A. S. Fraser is Professor Emeritus, Department of Statistics,
University of Toronto,
Toronto, Canada M5S 3G3 and Department of Statistical and Actuarial Sciences,
University of Western Ontario, London, Ontario, Canada N6A 5B7
\printead{e1}.}

\end{aug}

%
\vspace*{-8pt}
\begin{abstract}
Bayes [\textit{Philos. Trans. R. Soc. Lond.} \textbf{53} (1763)
370--418; \textbf{54} 296--325] introduced the observed likelihood function to statistical
inference and provided a weight function to calibrate the parameter;
he also introduced a~confidence distribution on the parameter space
but did not provide present justifications. Of
course the names likelihood and confidence did not appear until much later:
Fisher [\textit{Philos. Trans. R. Soc. Lond. Ser. A Math. Phys. Eng.
Sci.} \textbf{222} (1922) 309--368]
for likelihood and Neyman [\textit{Philos. Trans. R.~Soc. Lond. Ser. A
Math. Phys. Eng. Sci.} \textbf{237} (1937) 333--380] for confidence.
Lindley [\textit{J.~Roy. Statist. Soc. Ser. B} \textbf{20} (1958)
102--107] showed that the Bayes and the confidence
results were different when the model was not location.
This paper examines the occurrence of true statements from the Bayes
approach and from the confidence approach, and shows that the
proportion of true statements in the Bayes case depends
critically on the presence of linearity in the model; and
with departure from this
linearity the Bayes approach can be a poor approximation and be
seriously misleading.
Bayesian integration of weighted likelihood thus provides
a first-order linear approximation to confidence, but without
linearity can give substantially incorrect results.
\end{abstract}

%

\end{frontmatter}

\section{Introduction}

Statistical inference based on the observed likelihood function was
initiated by Bayes (\citeyear{Bay63}). This was, however, without the naming of the
likelihood function or the apparent recognition that likelihood
$L^0(\theta)=f(y^0; \theta)$ directly records\vadjust{\goodbreak} the amount of
probability at an observed data point $y^0$; such appeared much later
(Fisher, \citeyear{FisN1}).

Bayes' proposal applies directly to a model with translation invariance
that in current notation would be written $f(y-\theta)$; it recommended
that a weight function or mathematical prior $\pi(\theta)$ be applied
to the likelihood $L(\theta)$, and that the product
$\pi(\theta)L(\theta)$ be treated as if it were a joint density for
$(\theta, y)$. Then with observed data $y^0$ and the use of the
conditional probability lemma, a posterior distribution
$\pi(\theta|y)=c\pi(\theta)L^0(\theta)$ was obtained; this was viewed
as a description of possible values for $\theta$ in the presence of
data $y=y^0$. For the location model, as examined by the Bayes
approach, translation invariance suggests a~constant or flat prior
$\pi(\theta)=c$ which leads to the posterior distribution
$\pi(\theta|y^0)=f(y^0-\theta)$ and, in the scalar case, gives the
posterior survival probability $s(\theta) =\int^\infty_\theta
f(y^0-\alpha)\,d\alpha$, recording alleged probability to the right of a
value $\theta$.

The probability interpretation that would seemingly attach to this
conditional calculation is as follows: if the~$\theta$ values that
might have been present in the application can be viewed as coming from
the frequency pattern $\pi(\theta)$ with each $\theta$ value in turn
giving rise to a $y$ value in accord with the model and if the
resulting $y$ values that are close to $y^0$ are examined, then the
associated $\theta$ values have the pattern $\pi(\theta|y^0)$.

The complication is that $\pi(\theta)$ as proposed is a~ma\-thematical
construct and, correspondingly, $\pi(\theta|y^0)$ is just a
mathematical construct. The argument using the conditional probability
lemma does not produce probabilities from no probabilities: the
probability lemma when invoked for an application has two
distributions as input and one distribution as output; and it asserts
the descriptive validity of the output on the basis of the descriptive
validity of the two inputs; if one of the inputs is absent and an
artifact is substituted, then the lemma says nothing, and produces no
probabilities. Of course, other lemmas and other theory may offer
something appropriate.

We will see, however, that something different is readily available and
indeed available without the special translation invariance. We will
also see that the procedure of augmenting likelihood $L^0(\theta)$ with
a modulating factor that expresses model structure is a powerful first
step in exploring information contained in Fisher's likelihood
function.

An alternative to the Bayes proposal was introduced by Fisher (\citeyear{Fis30}) as
a confidence distribution. For the scalar-parameter case we can record
the percentage position of the data point $y^0$ in the distribution
having parameter value $\theta$,
\[
p(\theta)=p(\theta;y^0)=\int^{y^0}_{-\infty} f(y-\theta)\,dy.
\]
This records the proportion of the $\theta$ population that is less
than the value $y^0$. For a general data point~$y$ we have of course
that $p(\theta; y)$ is uniformly distributed on $(0, 1)$, and,
correspondingly, $p(\theta)$ from the data~$y^0$ gives the upper-tail
distribution function or survivor function for confidence, as
introduced by Fisher\break (\citeyear{FisN2}).~A~basic way of presenting confidence is
in terms of quantiles. If we set $p(\theta)=0.95$ and solve for~$\theta$,
we obtain $\theta=\hat\theta_{0.95}$ which is the value with
right tail confidence $95\%$ and left tail confidence $5\%$; this would
typically be called the $95\%$ lower confidence bound, and $(
\hat\theta_{0.95}, \infty)$ would be the corresponding $95\%$
confidence interval.

For two-sided confidence the situation has some subtleties that are
often overlooked. With the large data sets that have come from the
colliders of High Energy Physics, a Poisson count can have a mean at a
background count level or at a larger value if some proposed particle
is actually present. A common practice in the High Energy Physics
literature (Mandelkern, \citeyear{Man02}) has been to form two-sided confidence
intervals and to allow the confidence contributions in the two tails to
be different, thereby accommodating some optimality criterion; see
also some discussion in Section \ref{sec4}. In practice, this meant\vadjust{\goodbreak} that the
confidence lower bound shied away from the critical parameter lower
bound describing just the background radiation. This mismanaged the
detection of a new particle. Accordingly, our view is that two-sided
intervals should typically have equal or certainly designated amounts
of confidence in the two tails. With this in mind, we now restrict
the discussion to the analysis of the confidence bounds as described
in the preceding paragraph and view confidence intervals as being
properly built on individual confidence bounds with designated
confidence values.

\begin{figure}
\begin{tabular}{@{}c@{}}

\includegraphics{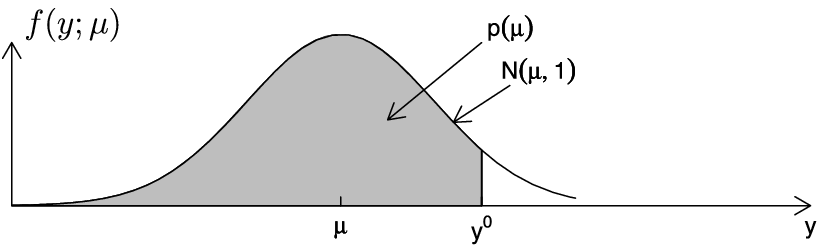}
\\
(a)\\

\includegraphics{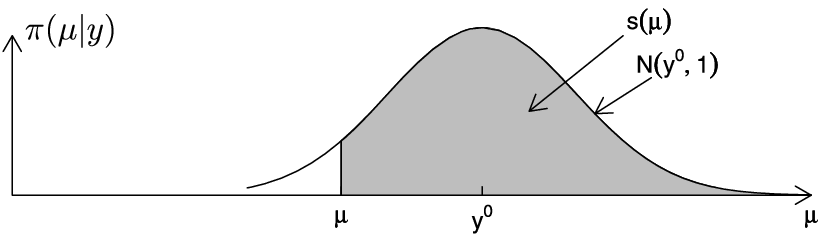}
\\
(b)
\end{tabular}
\caption{$\operatorname{Normal}(\mu, 1)$ model: The density of $y$ given
$\mu$
in \textup{(a)}; the posterior density of $\mu$ given $y^0$ in \textup{(b)}. The
$p$-value $p(\mu)$ from \textup{(a)} is equal to the survivor value $s(\mu)$ in
\textup{(b)}.} \label{normal}
\vspace*{3pt}
\end{figure}

As a simple example consider the $\operatorname{Normal}(\mu, \sigma_0^2)$, and let
$\phi(z)$ and $\Phi(z)$ be the standard normal density and
distribution functions. The $p$-value from data $y^0$ is
\[
p(\mu) =\int_{-\infty}^{y^0}\phi\biggl(\frac{y-\mu}{\sigma_0}\biggr)\,
dy= \Phi\biggl(\frac{y^0 - \mu}{\sigma_0}\biggr),
\]
which has normal distribution function shape dropping from $1$ at
$-\infty$ to $0$ at $+\infty$; it records the probability position of
the data with respect to a possible parameter value $\mu$; see Figure
\ref{normal}(a). From the confidence viewpoint, $p(\mu)$ is recording
the right tail confidence distribution function, and the confidence
distribution is $\operatorname{Normal} (y^0, \sigma^2_0)$.

%

The Bayes posterior distribution for $\mu$ using the invariant prior
has density $c\phi\{(y^0-\mu)/{\sigma_0}\}$; this is $\operatorname{Normal} (y^0,
\sigma^2_0)$. The resulting posterior survivor fun\-ction value is
\[
s(\mu)=\int_\mu^\infty
\phi\biggl(\frac{y^0-\alpha}{\sigma_0}\biggr)\,d\alpha=\Phi\biggl(\frac{y^0
- \mu}{\sigma_0}\biggr)
\]
and its values are indicated in Figure \ref{normal}(b); the function
provides a probability-type evaluation of the right tail interval
$(\mu, \infty)$ for the parameter. For this we have used the letter $s$
to suggest the ``survivor'' aspect of the Bayes analogue of the present
one-sided frequentist $p$-value.

\begin{figure}
\begin{tabular}{@{}c@{}}

\includegraphics{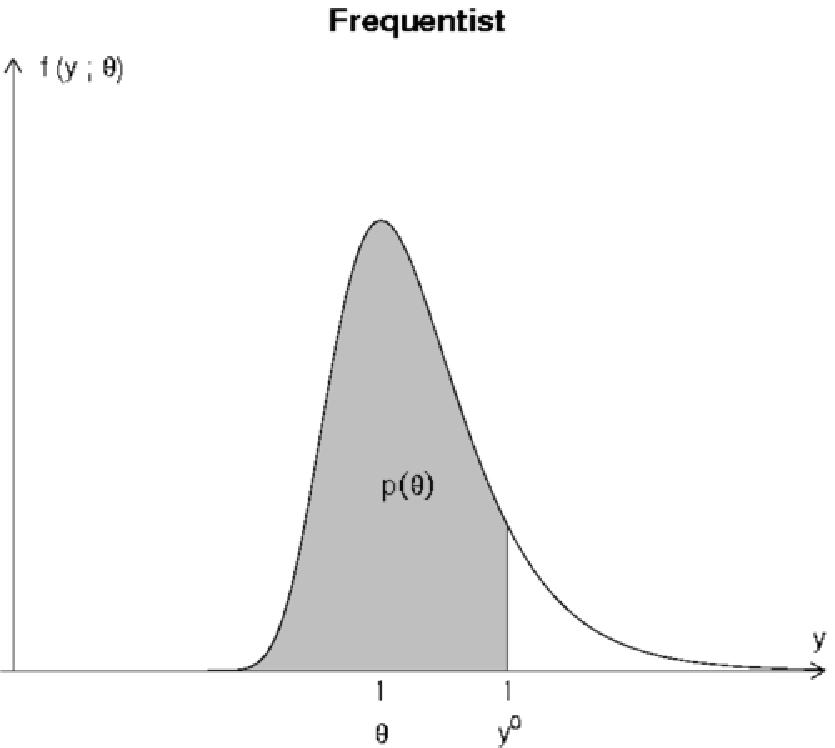}
\\
(a)\\

\includegraphics{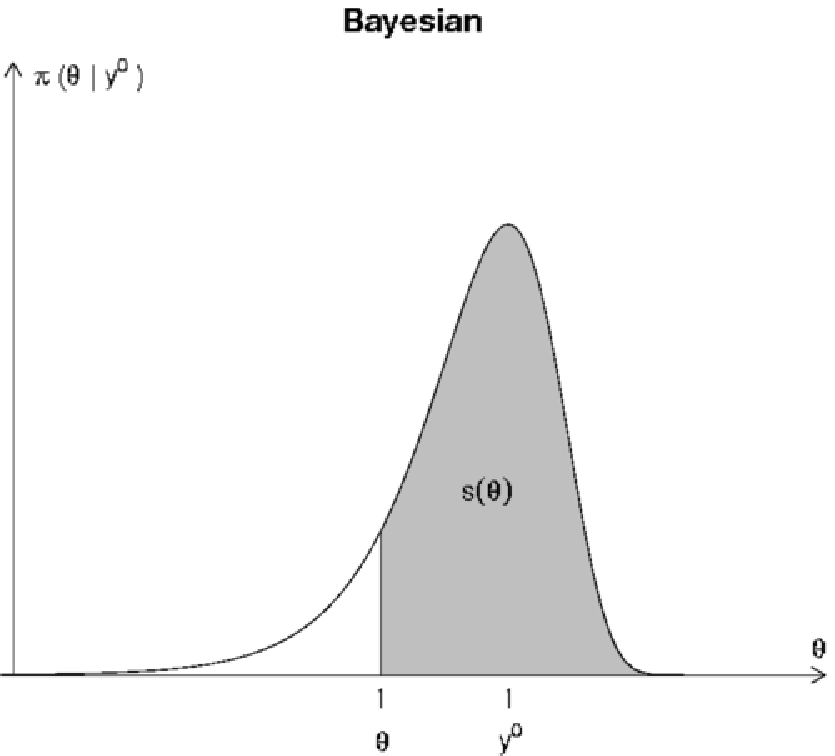}
\\
(b)
\end{tabular}
\caption{The extreme value $\operatorname{EV}(\theta,1)$ model: the density of~$y$
given $\theta$ in \textup{(a)}; the posterior density of $\theta$ given
$y^0$ in \textup{(b)}. The $p$-value $p(\theta)$ from \textup{(a)} is equal to the
survivor value $s(\theta)$ in~\textup{(b)}.} \label{extreme}
\end{figure}

For a second example consider the model \mbox{$y=\theta+z$}, where $z$ has
the standard extreme value distribution with density $g(z) = e^{-z}\exp\{-e^{-z}\}$ and distribution function $G(z)=\exp(-e^{-z})$. The
$p$-value from data~$y^0$ is
\begin{eqnarray*}
p(\theta)&=&\int_{-\infty}^{y^0}g(y-\theta)\,dy=G(y^0-\theta)\\&=&\exp
\bigl\{-e^{-(y^0-\theta)}\bigr\},
\end{eqnarray*}
which records the probability position of the data in the $\theta$
distribution; it can be viewed as a right tail distribution function
for confidence. The posterior distribution for $\theta$ using the Bayes
invariant prior has density $g(y^0-\theta)$ and can be described as a
reversed extreme value distribution centered at $y^0$. The posterior
survivor function is
\[
s(\theta)=\int_\theta^\infty g(y^0-\alpha)\,d\alpha=\exp
\bigl\{-e^{-(y^0-\theta)}\bigr\},
\]
and again agrees with the $p$-value $p(\theta)$; see Figure~\ref{extreme}.

Of course, in general for a location model $f(y-\theta)$ as examined
by Bayes, we have
\begin{eqnarray*}
p(\theta)&=&\int^{y^0}_{-\infty} f(y-\theta)\,dy =
\int^{y^0-\theta}_{-\infty} f(z)\,dz\\
& =& \int^\infty_\theta
f(y^0-\alpha)\,d\alpha= s(\theta),
\end{eqnarray*}
and, thus, the Bayes posterior distribution is equal to the confidence
distribution. Or, more directly, the Bayes posterior distribution is
just standard confidence.

Lindley (\citeyear{Lin58}) presented this result and under suitable change of
variable and parameter showed more: that the $p$-value and $s$-value
are equal if and only if the model $f(y; \theta)$ is a location model
$f(y-\theta)$. In his perspective then, this argued that the confidence
approach was flawed, confidence as obtained by inverting the $p$-value
function as a pivot. From a~different perspective, however, it argues
equally that the Bayes approach is flawed, and does not have the
support of the confidence interpretation unless the model is location.

Lindley objected also to the term probability being attached to the
original Fisher word for confidence, viewing probability as appropriate
only in reference to the conditional type calculations used by Bayes.
By contrast, repetition properties for confidence had been clarified by
Neyman (\citeyear{Ney37}). As a~consequence, in the discipline of statistics, the
terms probability and distribution were then typically not used in
the confidence context, but were in the Bayes context. The repetition
properties, however, do not extend to the Bayes calculation except for
simple location cases, as we will see; but they do extend for the
confidence inversion. We take this as strong argument that the term
probability is less appropriate in the Bayesian weighted likelihood
context than in the confidence inversion context.\looseness=1

The location model, however, is extremely special in that the parameter
has a fundamental linearity and this linearity is expressed in the use
of the flat prior with respect to the location parameter. Many
extensions of the Bayes mathematical prior have been proposed trying to
achieve the favorable behavior of the original Bayes, for example,
Jeffreys (\citeyear{Jef83}, \citeyear{Jef46}) and Bernardo (\citeyear{Ber79}). We refer to such priors as
default priors, priors to elicit information from an observed
likelihood function. And we will show that if the parameter departs
from a~basic linearity, then the Bayes posterior can be seriously
misleading. Specifically, we will show that with moderate departures
from linearity the Bayes calculation can give an acceptable
approximation to confidence, but that with more extreme departure from
linearity or with large parameter dimension it can give unacceptable
approximations.

John Tukey actively promoted a wealth of simple statistical methods as
a means to explore data; he referred to them as quick and dirty
methods. They were certainly quick using medians and ranges and other
easily accessible characteristics of data. And they were dirty in the
sense of ignoring characteristics that in the then currently correct
view were considered important. We argue that Bayes posterior
calculations can appropriately be called quick and dirty, quick and
dirty confidence.

There are also extensions of the Bayes approach allowing the prior to
reflect the viewpoint or judgment or prejudice of an investigator; or
to reflect the elicited considerations of those familiar with the
context being investigated. Arguments have been given that such a~viewpoint or consideration can be expressed as probability; but the
examples that we present suggest otherwise.

There are of course contexts where the true value~of the parameter
has come from a source with a~known distribution; in such cases the
prior is real, it is objective, and could reasonably be considered to
be a~part of an enlarged model. Then whether to include the prior
becomes a modeling issue. Also, in particular contexts, there may be
legal, ethical or moral issues as to whether such outside information
can be included. If included, the enlarged model is a probability model
and accordingly is not statistical: as such, it has no statistical
parameters in the technical sense and thus predates Bayes and can be
viewed as being probability itself not Bayesian. Why this would
commonly be included in the Bayesian domain is not clear; it is not
indicated in the original Bayes, although it was an area neglected by
the frequentist approach. Such a prior describing a~known source is
clearly objective and can properly be called an objective prior; this
conflicts, however, with some recent Bayesian usage where the term
objective is misapplied and refers to the mathematical priors that we
are calling default priors.

In Section \ref{sec2} we consider the scalar variable scalar parameter case and
determine the default prior that gives posteriors with reliable
quantiles; some details for the vector parameter case are also
discussed. In Section \ref{sec3} we argue that the only satisfactory way to
assess distributions for unobserved quantities is by means of the
quantiles of such distributions; this provides the basis then for
comparing the Bayesian and frequentist approaches.

In Sections \ref{sec4}--\ref{sec6} we explore a succession of examples that examine
how curvature in the model or in the parameter of interest can destroy
any confidence reliability in the default Bayes approach, and thus in
the Bayesian use of just likelihood to present a~distribution
purporting to describe an unknown parameter value.

In Sections \ref{sec7} and \ref{sec8} we discuss the merits of the conditional
probability formula when used with a~ma\-thematical prior and also the
merits of the optimality approach; then Section \ref{sec9} records a brief
discussion and Section \ref{sec10} a summary.

\section{But if the Model is Nonlinear}\label{sec2}

With a location model the confidence approach gives $p(\theta)$ and the
default Bayes approach gives $s(\theta)$, and these are equal. Now
consider things more generally and initially examine just a
statistical\vadjust{\goodbreak} mo\-del~$f(y;\theta)$ where both $y$ and $\theta$ are
scalar or real valued as opposed to vector valued, but without the
assumed linear relationship just discussed.

Confidence is obtained from the observed distribution function
$F^0(\theta)$ and a posterior is obtained from the observed density
function $f^0(\theta)$. For convenience we assume minimum continuity
and that $F(y;\theta)$ is stochastically increasing and attains
both~$0$ and $1$ under variation in $y$ or $\theta$. The confidence
$p$-va\-lue is directly the observed distribution function,
\[
p(\theta)=F^0(\theta)=F(y^0; \theta),
\]
which can be rewritten mechanically as
\[
p(\theta)=\int_{\theta}^{\infty} -F_{;\theta}(y^0; \alpha)\,
d\alpha;
\]
the subscript denotes partial differentiation with respect to the
corresponding argument. The default Bayes $s$-value is obtained from
likelihood, which is the observed density function $f(y^0; \theta) =
F_y(y^0; \theta)$:
\[
s(\theta)=\int_{\theta}^{\infty}\pi(\alpha)F_y(y^0; \alpha)\,
d\alpha.
\]

If $p(\theta)$ and $s(\theta)$ are in agreement, then the direct
comparison of the integrals implies that
\[
\pi(\theta)= \frac{-F_{;\theta}(y^0; \theta)}{F_y(y^0;\theta)}.
\]
This presents $\pi(\theta)$ as a possibly data-dependent\break prior.~Of
course, data dependent priors have a long but rather infrequent
presence, for example, Box and Cox (\citeyear{BoxCox64}), Wasserman (\citeyear{Was00}) and Fraser
et al. (\citeyear{Fraetal2010b}). The preceding expression for the prior can be
rewritten as
\[
\pi(\theta) = \frac{dy}{d\theta}\bigg|_{y^0}
\]
by directly differentiating the quantile function $y=y(u, \theta)$
for fixed $p$-value to $u = F(y; \theta)$ and taking the observed value, or by taking the total
differential of $F(y; \theta)$; Lindley's (\citeyear{Lin58}) result follows by
noting that the differential equation $dy/d\theta =a(\theta)/b(y)$ integrates to give a location model.

Now suppose we go beyond the simple case of the scalar model and allow
that $y$ is a vector of length $n$ and $\theta$ is a vector of length
$p$. In many applications $n>p$; but here we assume that dim $y$ has
been reduced to~$p$ by conditioning (see, e.g., Fraser, Fraser and
Staicu, \citeyear{FraFraSta2010c}), and that a smooth pivot $z(y, \theta)$ with density
$g(z)$ describes how the parameter affects the distribution of the
variable. The density for $y$ is available by inverting from pivot to
sample space:
\[
g(z)dz=f(y; \theta)\,dy = g\{z(y;\theta)\} |z_y(y; \theta)|\,dy,
\]
where the subscript again denotes partial differentiation.

For confidence a differential element is obtained by inverting from
pivot to parameter space:
\[
g(z)dz= g\{z(y^0; \theta)\} |z_{;\theta}(y^0; \theta)|\,d\theta.
\]
And for posterior probability the differential element is obtained as
weighted likelihood
\[
g(z)dz = g\{z(y^0; \theta)\} |z_y(y^0; \theta)| \pi(\theta)\,d\theta.
\]
The confidence and posterior differential elements are equal if
\[
\pi(\theta) = \frac{|z_{;\theta}(y^0; \theta)|}{|z_y(y^0;
\theta)|};
\]
we call this the default prior for the model $f(y;\theta)$ with data
$y^0$. As $dy/d\theta= z_y^{-1}(y^0; \theta)z_{;\theta}(y^0; \theta)$
for fi\-xed~$z$, we will have confidence equal to posterior if
$\pi(\theta) = |dy/d\theta|_{y^0}$, a simple extension of the
scalar case. The matrix ${dy} / {d\theta} |_{y^0}$ can be called the
sensitivity of the parameter at the data point $y^0$ and the
determinant provides a natural weighting or scaling function
$\pi(\theta)$ for the parameter; this sensitivity is just presenting
how parameter change affects the model and is recording this just at
the relevant point, the observed data.

\section{How to Evaluate a Posterior Distribution}\label{sec3}

(i) \textit{Distribution function or quantile function}. In the scalar
parameter case, both $p(\theta)$ and $s(\theta)$ have the form of a
right tail distribution function or survivor function. In the Bayesian
framework, the function $s(\theta)$ is viewed as a distribution of
posterior probability. In the frequentist framework, the function
$p(\theta)$ can be viewed as a distribution of confidence, as
introduced by Fisher (\citeyear{Fis30}) but originally called fiducial; it has
long been a familiar theme, frequentist or Bayesian, that it is inappropriate to treat such a
function as a distribution describing possible values for the true
parameter.

For a scalar parameter model with data, the Bayes and the confidence
approaches with data each lead to a probability-type evaluation on the
parameter space; and these can be different as Lindley (\citeyear{Lin58})
demonstrated and as we have quantified\vadjust{\goodbreak} in the preceding section. Surely
then, they both cannot be correct. So, how to evaluate such posterior
distributions for the parameter?

A probability description is a rather complex thing even for a scalar
parameter: ascribing a probability-type assessment to one-sided
intervals, two-sided intervals, and more general sets. What seems more
tangible but, indeed, is equivalent is to focus on the reverse, the
quantiles: choose an amount $\beta$ of probability and then determine
the corresponding quantile $\hat\theta_\beta$, a value with the
alleged probability $1-\beta$ to the left and with $\beta$ to the
right. We then have that a~particular interval ($\hat\theta_\beta,
\infty$) from the data has the alleged amount $\beta$. Here we focus
on such quantiles~$\hat\theta_\beta$ on the scale for~$\theta$. In
particular, we might examine the $95\%$ quantile $\hat\theta_{0.95}$,
the median quantile~$\hat\theta_{0.50}$, the $5\%$ quantile
$\hat\theta_{0.05}$, and others, all as part of examining an alleged
distribution for~$\theta$ obtained from the data.

For the $\operatorname{Normal}(\mu, \sigma_0^2)$ example with data $y^0$, the
confidence approach gives the $\beta$-level quantile
\[
\hat\mu_\beta= y^0 - z_\beta\sigma_0,
\]
where $\Phi(z_\beta) = \beta$ as based on the standard normal
distribution function $\Phi$. In particular, the $95\%,  50\%$ and
$5\%$ quantiles are
\begin{eqnarray*}
\hat\mu_{0.95} &=& y^0- 1.64\sigma_0,\quad \hat\mu_{0.50} = y^0,\\
\hat\mu_{0.05} &=& y^0+ 1.64\sigma_0;
\end{eqnarray*}
and the corresponding confidence intervals are
\[
(y^0-1.64\sigma_0, \infty),\quad (y^0, \infty),\quad
(y^0+1.64\sigma_0, \infty),
\]
with the lower confidence bound in each case recording the
corresponding quantile.

%
\begin{figure}

\includegraphics{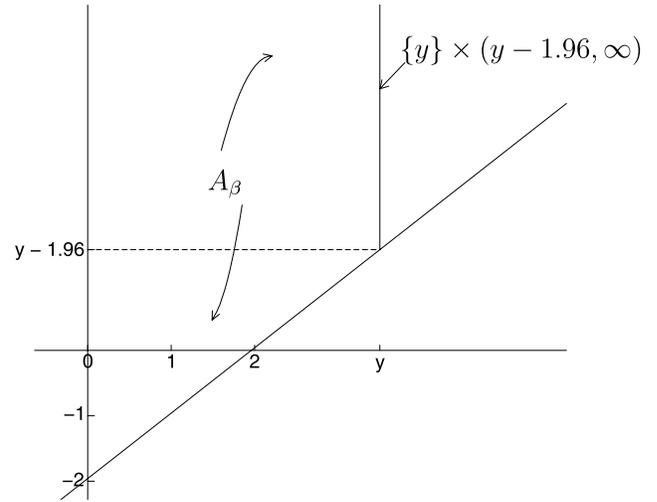}

\caption{The $97.5\%$ allegation for the $\operatorname{Normal}$ confidence
procedure, on the $(y, \theta)$-space.} \label{confidence}
\vspace*{3pt}
\end{figure}

Now more generally suppose we have a model $f(y;\allowbreak\theta)$ and data
$y^0$, and that we want to evaluate a proposed procedure, Bayes,
frequentist or other, that~gi\-ves a probability-type evaluation of where
the true parameter $\theta$ might be. As just discussed, we can focus
on some level, say, $\beta$, and then examine the corresponding
quantile $\hat\theta_\beta$ or the related interval ($\hat\theta
_\beta, \infty$). In any particular instance, either the true $\theta$ is in
the interval ($\hat\theta_\beta, \infty$), or it is not. And yet the
procedure has put forward a numerical level $\beta$ for the presence of
$\theta$ in ($\hat\theta_\beta, \infty$). What does the asserted level
$\beta$ mean?

(ii) \textit{Evaluating a proposed quantile}. The definitive evaluation
procedure is in the literature: use a Neyman (\citeyear{Ney37}) diagram. The model
$f(y; \theta)$ sits on the space $S \times\Omega$ which here is the
real line for $S$ \textit{cross} the real line for $\Omega$; this is just
the plane $R^2$. For any particular $y$ the procedure gives a parameter
interval\vadjust{\goodbreak} ($\hat\theta_\beta(y), \infty$); if we then assemble the
resulting intervals, we obtain a~region
\[
A_\beta= \bigcup\{y\} \times(\hat\theta(y), \infty) =\{(y, \theta)\dvtx
\theta  \mbox{ in }   (\hat\theta(y) , \infty)\}
\]
on the plane. For the confidence procedure in the simple $\operatorname{Normal}(\theta,1)$ case,
Figure \ref{confidence} illustrates the $97.5\%$
quantile $\hat\theta_{0.975}$ for that confidence procedure; the region
$A_\beta= A_{0.975}$ is to the upper left of the angled line and it
represents the $\beta=97.5\%$ allegation concerning the true $\theta$,
as proceeding from the confidence procedure.

Now, more generally for a scalar parameter, we suggest that the sets
$A_\beta$ present precisely the essence of a~posterior procedure: how
the procedure presents information concerning the unknown $\theta$
value. We can certainly examine these for various values of $\beta$ and
thus investigate the merits of any claim implicit in the alleged levels
$\beta$.

The level $\beta$ is attached to the claim that $\theta$ is in
($\hat\theta_\beta(y), \infty$), or, equivalently, that ($y, \theta$)
is in the set~$A_\beta$. In any particular instance, there is of course
a true value $\theta$, and either it is in \{$\hat\theta_\beta(y),
\infty$\} or it is not in \{$\hat\theta_\beta(y), \infty$\}. And the
true $\theta$ did precede the generation of the observed $y$ in full
accord with the probabilities given by the model. Accordingly, a~value
$\theta$ for the parameter in the model implies an actual Proportion
of true assertions consequent to that $\theta$ value:
\[
\operatorname{Propn} (A_\beta; \theta)= \operatorname{Pr}\{A_\beta \mbox{ includes }  (y, \theta); \theta\}.
\]
This allows us to check what relationship the actual Proportion bears
to the value $\beta$ asserted by the procedure: is it really $\beta$ or
is it something else?\vadjust{\goodbreak}

Of course, there may be contexts where in addition we have that the
$\theta$ value has been generated by some random source described by an
available prior density $\pi(\theta)$, and we would be interested in
the associated Proportion,
\[
\operatorname{Propn} (A_\beta; \pi)=\int\operatorname{Propn} ( A_\beta; \theta)
\pi(\theta)\,d\theta,
\]
presenting the average relative to the source densi\-ty~$\pi(\theta)$.

(iii) \textit{Comparing proposed quantiles}. For the Bayes procedure with
the special original linear model\break \mbox{$f(y-\theta)$} we have by the usual
calculations that
\[
\operatorname{Propn} (A_\beta; \theta) \equiv\beta
\]
for all $\theta$ and $\beta$: the alleged level $\beta$ agrees with the
actual Proportion of true assertions that are made. And, more
generally, if the $\theta$ value has been generated by a source
$\pi(\theta)$, then it follows that the alleged level $\beta$ does
agree with the actual Proportion: thus, $\operatorname{Propn} (A_\beta; \pi)
\equiv\beta$.

For the standard confidence procedure in the context of an arbitrary
continuous scalar model $f(y;\theta)$, we have by the standard
calculations that
\begin{eqnarray*}
\operatorname{Propn} (A_\beta; \theta) &\equiv&\operatorname{Pr} \{(y, \theta) \mbox{ in }
 A_\beta; \theta\}\\
  &\equiv&\operatorname{Pr} \{F(y; \theta) \leq\beta;
\theta\} \equiv\beta
\end{eqnarray*}
for all $\theta$ and $\beta$. Of course, in the special Bayes location
model $f(y-\theta)$ the Bayes original procedure does coincide with the
confidence procedure: the original Bayes was confidence in disguise.

Now for some proposed procedure having a region~$A_\beta$ with alleged
level $\beta$, there is of course the possibility that the actual
Proportion is less than $\beta$ for some~$\theta$ and is greater than
$\beta$ for some other $\theta$ and yet when averaged with a particular
prior $\pi(\theta)$ gives a revised $\operatorname{Propn} (A_\beta; \pi)$ that
does have the value $\beta$; the importance or otherwise of this we
will discuss later.

But we now ask, what is the actual Proportion for a~Bayes procedure in
nonlocation models? Toward this, we next examine a succession of
examples where the linearity is absent to varying degrees, where the
parameter to variable relationship is nonlinear!

\section{Nonlinearity and Bounded Parameter: The Errors are $O(1)$}\label{sec4}

We first examine an extreme form of nonlinearity, where the range of
the parameter is bounded. This is a~familiar\vadjust{\goodbreak} problem in the current
High Energy Physics of particle accelerators and the related
search and detection of possible new particles: a particle count
has a~$\operatorname{Poisson} (\theta)$ distribution but $\theta$ is bounded
below by~$\theta_0$, which represents the contribution from background
radiation. For some discussion see Mandelkern (\citeyear{Man02}), Reid and Fraser
(\citeyear{ReiFra03}) and Fraser, Reid and Wong (\citeyear{FraReiWon04}).

The critical issues are more easily examined in a continuous context.
For this, suppose that $y$ is $\operatorname{Normal}(\theta, \sigma_0^2$) and that it
is known that $\theta\geq\theta_0$ with an interest in detecting
whether $\theta$ is actually larger than $\theta_0$; let $y^0$ be the
observed data point; this continuous version was also mentioned in
Mandelkern (\citeyear{Man02}), Woodroofe and Wang (\citeyear{WooWan00}) and Zhang and Woodroofe
(\citeyear{ZhaWoo03}). For convenience here and without loss of generality, we take
the known $\sigma_0 = 1$ and the lower bound $\theta_0=0$.

%
\begin{figure}
\begin{tabular}{@{}c@{}}

\includegraphics{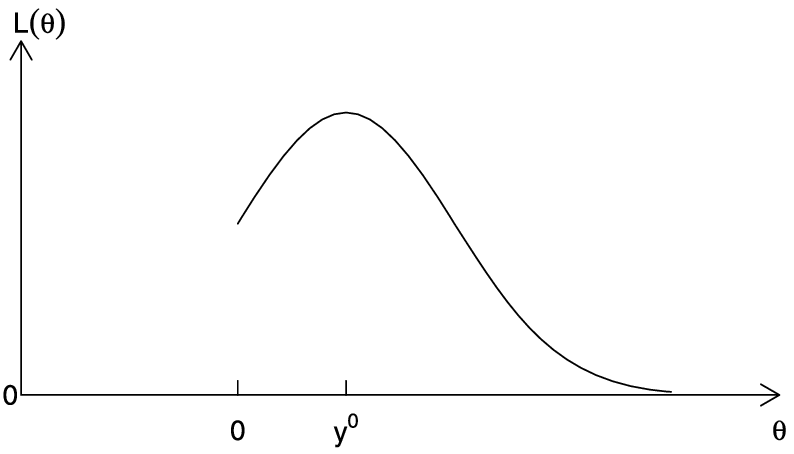}
\\
(a)\\

\includegraphics{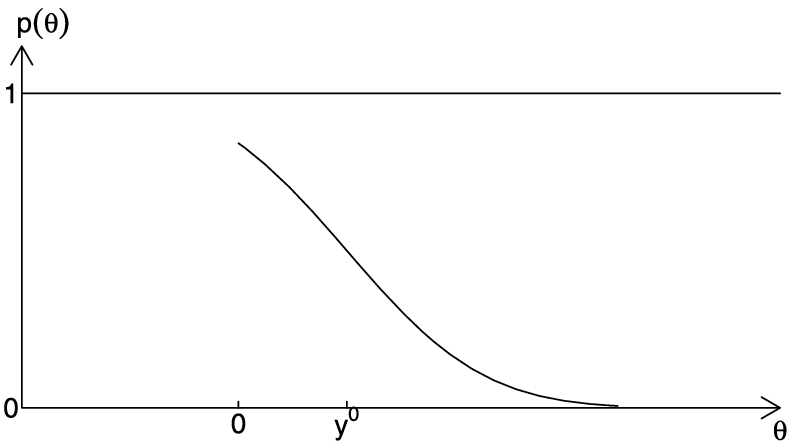}
\\
(b)\\

\includegraphics{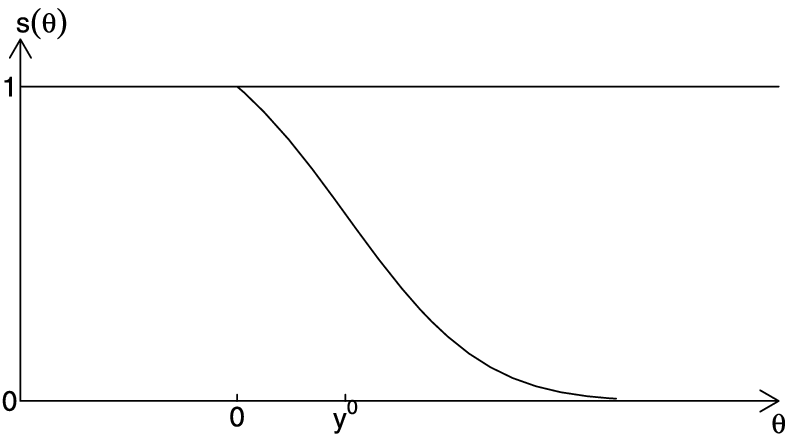}
\\
(c)
\end{tabular}
\caption{The $\operatorname{Normal} (\theta, 1)$ with $\theta\geq\theta_0 =
0$: \textup{(a)} the likelihood function $L(\theta)$; \textup{(b)} $p$-value function
$p(\theta)=\Phi(y^0-\theta)$; \textup{(c)} $s$-value function
$s(\theta)=\Phi(y^0-\theta) / \Phi(y^0)$.} \label{truncatedN}
\end{figure}

From a frequentist viewpoint, there is the likelihood
\[
L^0(\theta) = c \phi(y^0 - \theta)
\]
recording probability at the data, again using $\phi(z)$ for the
standard normal density. And also there is the $p$-value
\[
p(\theta)=\Phi(y^0 - \theta)
\]
recording probability left of the data. They each offer a basic
presentation of information concerning the parameter value $\theta$;
see Figure \ref{truncatedN}(a) and (b). Also note
that $p(\theta)$ does not reach the value $1$ at the lower limit
$\theta_0$ for~$\theta$; of course, the $p$-value is just recording
the statistical position of the data $y^0$ under possible~$\theta$
values, so there is no reason to want or expect such a limit.

First consider the confidence approach. The interval $(0,\beta)$ for
the $p$-value function gives the interval $\{\max (\theta_0, y^0
-z_\beta), \infty\}$ for $\theta$ when we acknowledge the lower bound,
or gives the interval $(y^0 -z_\beta, \infty)$ when we ignore the lower
bound. In either case the actual Proportion is equal to the alleged
value $\beta$, regardless of the true value of $\theta$. There might
perhaps be mild discomfort that if we ignore the lower bound and
calculate the interval, then it can include parameter values that are
not part of the problem; but nonetheless the alleged level is
valid.\looseness=1

Now consider the default Bayes approach. The model $f(y; \theta)=
\phi(y^0-\theta)$ is translation invariant for $\theta\geq\theta_0$,
and this would indicate the constant prior $\pi(\theta) = c$, at least
for $\theta\geq\theta_0$. Combining the prior and likelihood
and\vadjust{\goodbreak}
norming as usual gives the posterior density
\[
\pi(\theta|y^0) = \frac{\phi(y^0-\theta)}{\Phi(y^0)},\quad \theta
\geq0,
\]
and then gives the posterior survivor value
\[
s(\theta) = \frac{\Phi(y^0 - \theta)}{\Phi(y^0)},\quad \theta\geq0.
\]
See Figure \ref{truncatedN}(c). The $\beta$-quantile of this truncated
normal distribution for $\theta$ is obtained by setting\break \mbox{$s(\theta) = \beta$} and solving for $\theta$:
\[
\hat\theta_\beta= y^0 - z_{\beta\Phi(y^0)},
\]
where again $z_\gamma$ designates the standard normal
$\gamma$-quantile.

We are now in a position to calculate the actual Proportion, namely,
the proportion of cases where it is true that $\theta$ is in the
quantile interval, or, equivalently, the proportion of cases where
($\hat\theta_\beta, \infty$) includes the true $\theta$ value:
\begin{eqnarray*}
\operatorname{Propn}(\theta) &=& \operatorname{Pr} \bigl\{y-z_{\beta\Phi(y)} <
\theta\dvtx
\theta\bigr\}\\
&=& \operatorname{Pr} \bigl\{ z<z_{\beta\Phi(\theta+z)}\bigr\},
\end{eqnarray*}
where $z$ is taken as being $\operatorname{Normal}(0, 1)$; this expression can be
written as in integral $\int_S\phi(z)\,dz$ with $S=\{z\dvtx
\Phi(z)<\beta\Phi(\theta+z)\}$ and can routinely be evaluated
numerically for particular values of $\theta$ and $\beta$.
In particular, for $\theta$ at the lower limit $\theta=\theta_0=0$
the coverage set $S$ becomes $S=\{z\dvtx\Phi(z)<\beta\Phi(z)\}$, which
is clearly the empty set unless $\beta=1$.
In particular, at the lower limit $\theta=\theta_0=0$ the $
\operatorname{Propn}(\theta_0)$ has the phenomenal value zero, $
\operatorname{Propn}(\theta_0)=0$, which is a consequence of the empty set just
mentioned; certainly an unusual performance property for a claimed
Bayes coverage of, say, $ \beta$ when, as typical, $\beta$ is not zero.

%
\begin{figure}

\includegraphics{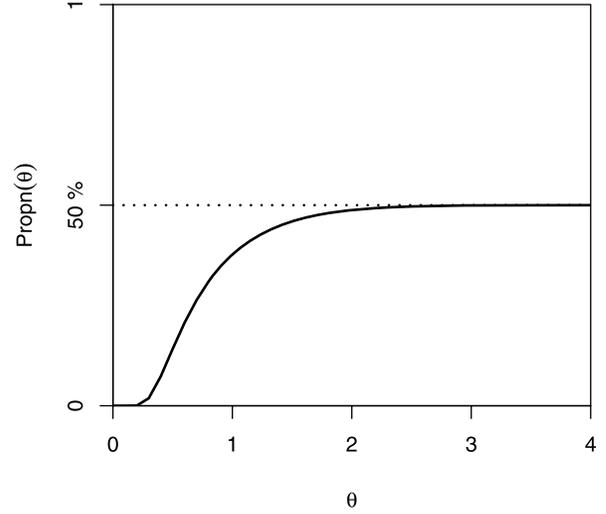}

\caption{$\operatorname{Normal}$ with bounded mean: the actual Proportion for the
claimed level $\beta=50\%$ is strictly less than the claimed $50\%$.}
\label{Ex1_1}
\end{figure}

%
\begin{figure}

\includegraphics{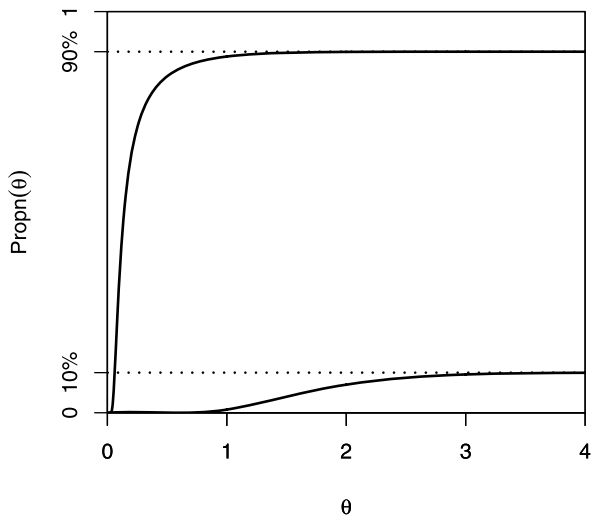}

\caption{$\operatorname{Normal}$ with bounded mean: the actual
Proportions for the claimed level $\beta=90\%$ and $\beta=10\%$ are
strictly less than the claimed.} \label{Ex1_2}
\end{figure}

In Figure \ref{Ex1_1} we plot this Proportion against $\beta$ for the
$\beta=50\%$ quantile; and we note that it is uniformly less than the
nominal, the claimed $50\%$. In particular, at the lower limit
$\theta=\theta_0=0$ the\break $\operatorname{Propn}(\theta_0)=0$ has the phenomenal
value zero, as mentioned in the preceding paragraph; certainly an
unusual performance property for a claimed Bayes coverage of $
\beta=50\%$! Then\vadjust{\goodbreak} in Figure \ref{Ex1_2} we plot the proportion for
$\beta=90\%$ and for $\beta=10\%$; again we note that the actual
Proportion is uniformly less than the claimed value, and again $
\operatorname{Propn}(\theta)$ has the extraordinary coverage value $0$ when the
parameter is at the lower bound $0$.
Of course, the departure would be in the other direction in the case of
an upper bound.

\begin{figure*}[t!]
\begin{tabular}{@{}c@{}}

\includegraphics{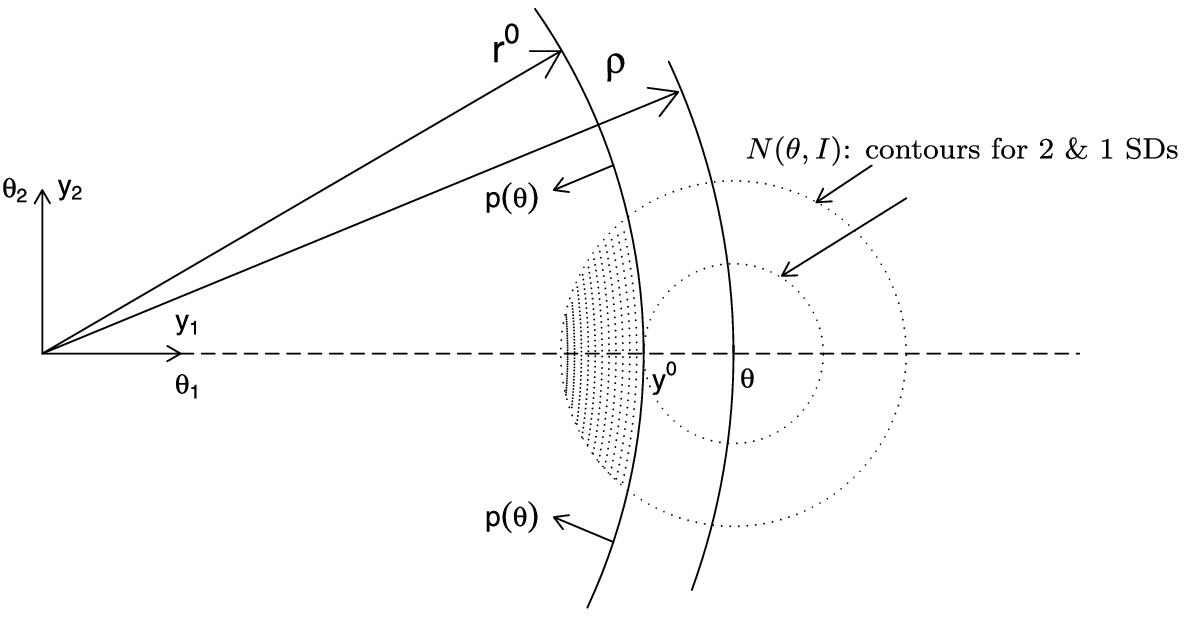}
\\
(a)\\

\includegraphics{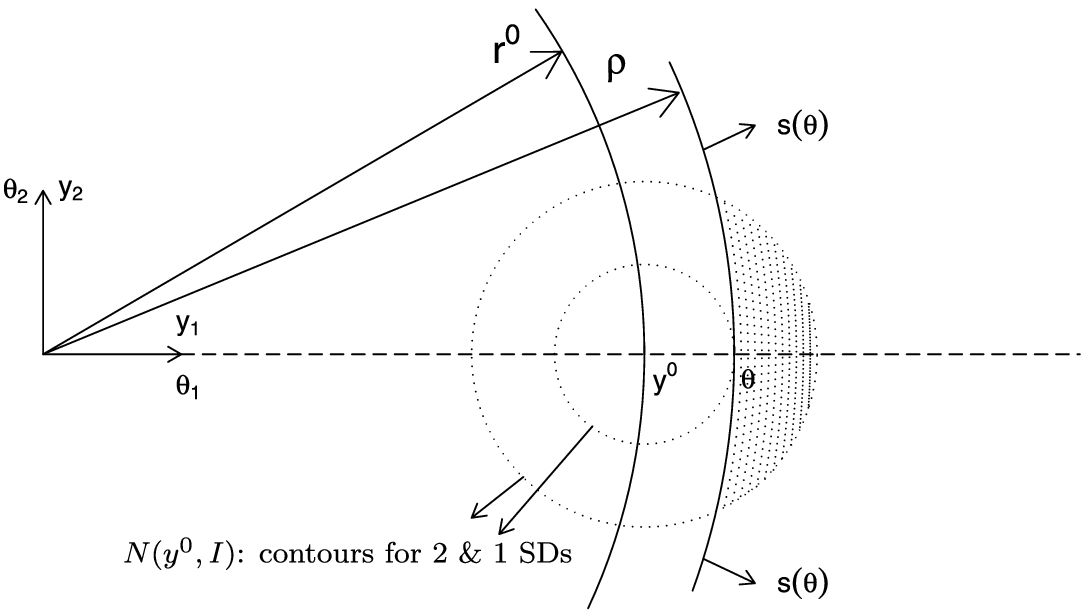}
\\
(b)
\end{tabular}
\caption{\textup{(a)} The model is $N(\theta; I)$; region for
$p(\theta)$
is shown. \textup{(b)} The posterior distribution for $\theta$ is $N(y^0; I)$;
region for $s(\theta)$ is shown.} \label{circle}
\vspace*{3pt}
\end{figure*}

In summary, in a context with a bound on the parameter, the performance
error with the
Bayes calculation can be of asymptotic order $O(1)$.

\section{\texorpdfstring{Nonlinearity and Parameter Curvature: The Errors are $O(\lowercase{n}^{-1/2})$}
{Nonlinearity and Parameter Curvature: The Errors are O(n^{-1/2})}}\label{sec5}

A bound on a parameter as just discussed is a~ra\-ther extreme form of
nonlinearity. Now consider a~ve\-ry direct and common form of curvature.
Let ($y_1, y_2$) be $\operatorname{Normal} (\theta; I)$ on $R^2$ and consider the
quadratic interest parameter $(\theta_1^2 + \theta_2^2)$, or the
equivalent $\rho(\theta) = {(\theta_1^2 + \theta_2^2)}^{1/2}$ which has
the same dimensional units as the $\theta_i$; and let $y^0 = (y^0_1,
y^0_2)$ be the observed data. For asymptotic analysis we would view
the present variables as being derived from some antecedent sample of
size $n$ and they would then have the $\operatorname{Normal} (\theta, I/\allowbreak n)$
distribution.

From the frequentist view there is an observable variable $r=(y_1^2 +
y_2^2)^{1/2}$ that in some pure physical sense measures the parameter
$\rho$. It has a noncentral chi distribution with noncentrality $\rho$
and degrees of freedom 2. For convenience we let $\chi_2(\rho)$
designate such a variable with distribution function $H_2(\chi, \rho)$,
which is typically available in computer packages; and its square can
be expressed as $\chi^2_2=(z_1 + \rho)^2 + z_2^2$ in terms of standard
normal variables and it has the noncentral chi-square distribution with
$2$ degrees of freedom and noncentrality usually described by
$\rho^2$. The distribution of $r$ is free of the nuisance parameter
which can conveniently be taken as the polar angle $\alpha= {\rm
arctan}(\theta_2/\theta_1)$. The resulting $p$-value function for
$\rho$ is
%
\begin{equation}\label{*}
p(\rho) =\operatorname{Pr}\{\chi_2(\rho) \leq r^0\} = H_2(r^0; \rho).
\end{equation}
See Figure \ref{circle}(a), where for illustration we examined the
behavior for $\theta= y^0 + 1$.

%
\begin{figure*}[t!]

\includegraphics{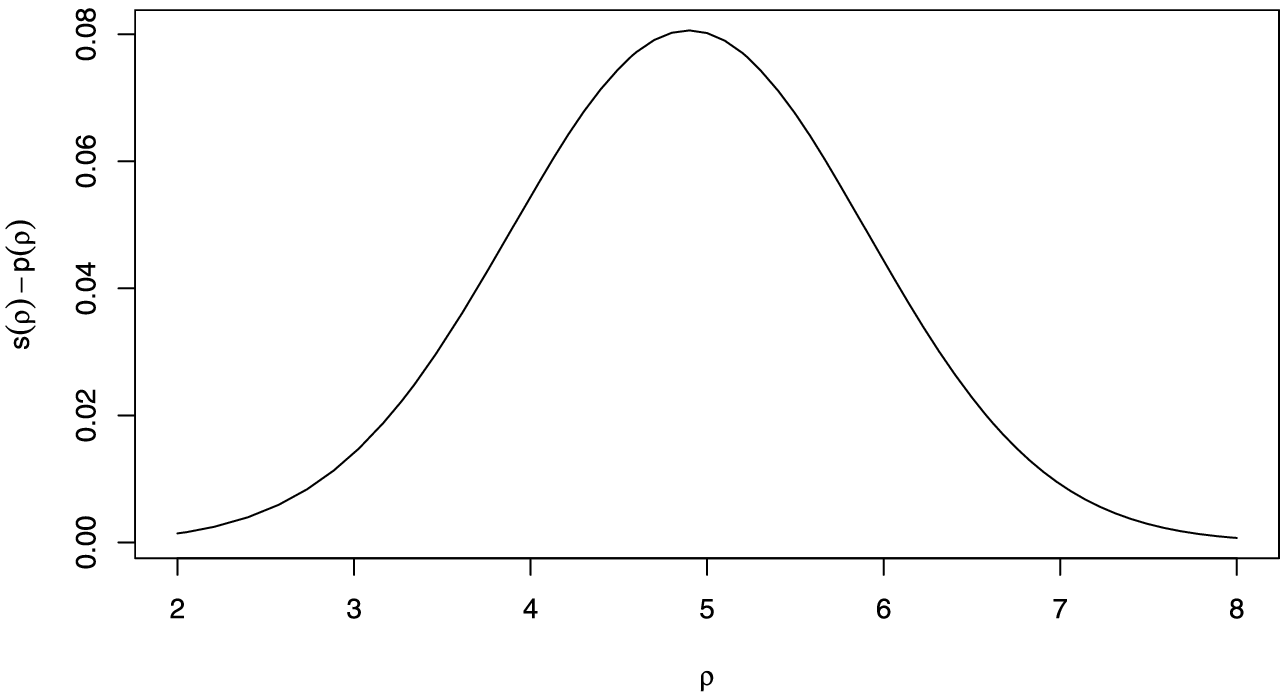}

\caption{The Bayes error $s(\rho)-p(\rho)$
from the $N(\theta, I)$ model with data $y^0=(5, 0)$.} \label{bayes
excess}
\end{figure*}

From the frequentist view there is the directly measured $p$-value
$p(\rho)$ with a $\operatorname{Uniform} (0,1)$ distribution, and any $\beta$ level
lower confidence quantile is available immediately by solving
$\beta=H_2(r^0; \rho)$ for $\rho$ in terms of $r^0$.

From the Bayes view there is a uniform prior\break \mbox{$\pi(\theta) = c$} as
directly indicated by Bayes (\citeyear{Bay63}) for a location model on the plane
$R^2$. The corresponding posterior distribution for $\theta$ is then
$\operatorname{Normal} (y^0; I)$ on the plane. And the resulting marginal posterior
for~$\rho$ is described by the generic variable $\chi_2(r^0)$. As~$r$
is stochastically increasing in $\rho$, we have that the Bayes analog
of the $p$-value is the posterior survivor value obtained by an upper
tail integration
%
\begin{equation}\label{**}
s(\rho) =\operatorname{Pr} \{ \chi_2(r^0) \geq\rho\} = 1 - H_2(\rho; r^0).
\end{equation}

The Bayes $s(\rho)$ and the frequentist $p(\rho)$ are actually quite
different, a direct consequence of the obvious curvature in the
parameter $\rho=(\theta_1^2 + \theta_2^2)^{1/2}$. The presence of the
difference is easily assessed visually in Figure \ref{circle} by noting
that in either case there is a rotationally symmetric normal
distribution with unit standard deviation which is at the distance
$d=1$ from the curved boundary used for the probability calculations,
but the curved boundary is cupped away from the $\operatorname{Normal}$ distribution in
the frequentist case and is cupped toward the $\operatorname{Normal}$ distribution in
the Bayes case; this difference is the direct source of the Bayes
error.

From (\ref{*}) and (\ref{**}) we can evaluate the posterior error
$s(\rho) - p(\rho)= 1 - H_2(\rho; r^0) - H_2(r^0; \rho) $ which is
plotted against $\rho$ in Figure \ref{bayes excess} for $r^0 = 5$. This
Bayes error here is always greater than zero. This happens widely with
a parameter that has curvature, with the error in one or other
direction depending on the curvature being positive or negative
relative to increasing values of the parameter. Some aspects of this
discrepancy are discussed in David, Stone and Zidek (\citeyear{DawStoZid73}) as a
marginalization paradox.

Now in more detail for this example, consider the~$\beta$ lower
quantile $\hat\rho_\beta$ of the Bayes posterior distribution for the
interest parameter $\rho$. This $\beta$ quantile for the parameter
$\rho$ is obtained from the $\chi_2(r^0)$ posterior distribution for
$\rho$ giving
\[
\hat\rho_\beta= \chi_{1-\beta}(r^0),
\]
where we now use $\chi_\gamma(r)$ for the $\gamma$ quantile of the
noncentral chi variable with 2 degrees of freedom and noncentrality
$r$, that is, $H_2(\chi_\gamma; r) = \gamma$. We are now in a~position
to evaluate the Bayes posterior proposal for $\rho$. For this let
$\operatorname{Propn}(A_\beta; \theta)$ be the proportion of true assertions that
$\rho$ is in $A_\beta\!=\!\{\hat\rho_\beta(r), \infty\}$; we have
\begin{eqnarray*}
\operatorname{Propn} (A_\beta; \rho) & = &  \operatorname{Pr} \{ \rho \mbox{ in }  (\hat
\rho_\beta(r), \infty); \rho\} \\
& = & \operatorname{Pr} \{\hat\rho_\beta(r) \leq\rho; \rho\} \\
& = & \operatorname{Pr} \{\chi_{1-\beta} (r) \leq\rho; \rho\},
\end{eqnarray*}
where the quantile $\hat\rho_\beta(r)$ is seen to be the $(1-\beta)$
point of a noncentral chi variable with degrees of freedom $2$ and
noncentrality $r$, and the noncentrality $r$ has a noncentral chi
distribution with noncentrality $\rho$. The actual Proportion under a
parameter value $\rho$ can thus be presented as
\begin{eqnarray*}
\operatorname{Propn} (A_\beta; \rho) & = & \operatorname{Pr} [\chi_{1-\beta} \{\chi
_2(\rho)\} \leq\rho; \rho] \\
& = & \operatorname{Pr} [1-\beta< H_2\{\rho; \chi_2(\rho)\}],
\end{eqnarray*}
which is available by numerical integration on the real line for any
chosen $\beta$ value.

%
\begin{figure}

\includegraphics{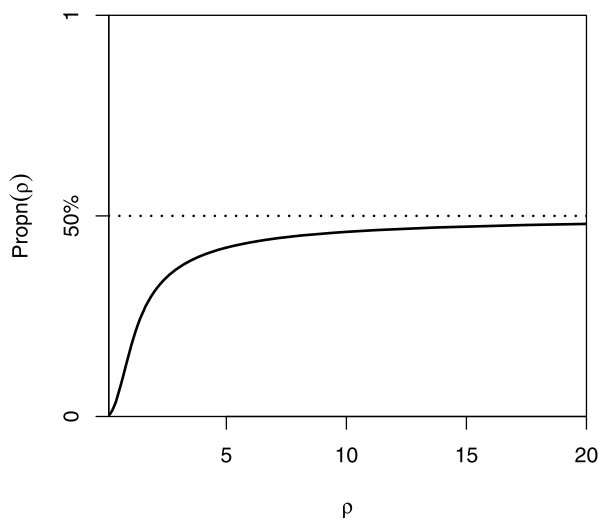}

\caption{Proportion with claimed level $\beta=50\%$.}
\label{Ex2_1}
\vspace*{3pt}
\end{figure}

%
\begin{figure}

\includegraphics{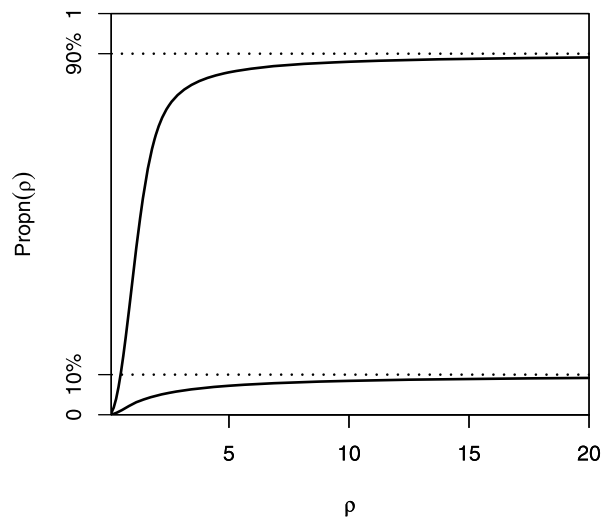}

\caption{Proportion for claimed $\beta=90\%$ and for
claimed $\beta=10\%$: strictly less than the claimed.} \label{Ex2_2}
\end{figure}

We plot the actual $\operatorname{Propn}(A_{50\%}; \rho)$ against $\rho$ in
Figure \ref{Ex2_1} and note that it is always less than the alleged
$50\%$. We then plot the Proportion for $\beta= 90\%$ and for $\beta=
10\%$ in Figure \ref{Ex2_2} against $\rho$, and note again that the
plots are always less than the claimed values $95\%$ and $5\%$. This
happens generally for all possible quantile levels $\beta$, that the
actual Proportion is less than the alleged probability. It happens for
any chosen value for the parameter; and it happens for any prior
average of such $\theta$ values. If by contrast the center of curvature
is to the right, then the actual Proportion is reversed and is larger
than the alleged.

In summary, in the vector parameter context with a curved interest
parameter the performance error with the Bayes calculation can be of
asymptotic order $O(n^{-1/2})$.

%
\begin{figure}

\includegraphics{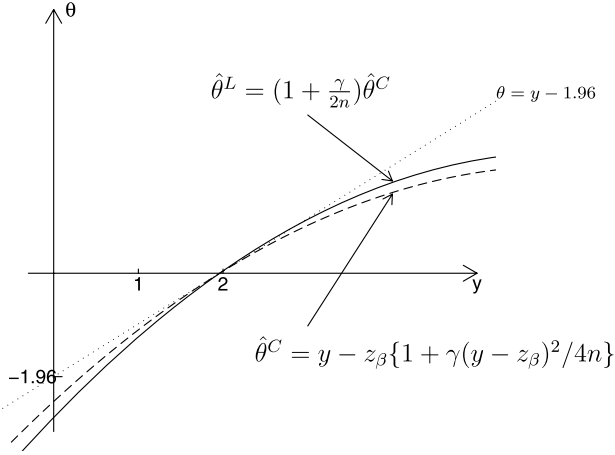}

\caption{The $97.5\%$ confidence quantile
$\hat\theta^C(y)= y-\allowbreak1.96\{1+\gamma(y- 1.96)^2 / 4n\}$. The $97.5\%$
likelihood quantile $\hat\theta^L(y)=(1+\frac{\gamma}{2n})[y-1.96\{1+\gamma(y-1.96)^2 / 4n\}]$ is a~vertical rescaling
about the origin; the $97.5\%$ Bayes quantile $\hat\theta^B(y)$ with
prior $\exp\{a/n + c\theta/n\}$ is a~vertical rescaling plus a
lift $a/n$ and a tilt $cy/n$. Can this prior lead to a confidence
presentation? No, unless the prior depends on the data or on the level
$\beta$.} \label{F}
\end{figure}

\section{\texorpdfstring{Nonlinearity and Model Curvature:  The Errors are $O(\lowercase{n}^{-1})$}
{Nonlinearity and Model Curvature:  The Errors are O(n^{-1})}}
\label{sec6}

(i) \textit{The model and confidence bound}. Taylor series expansions
provide a powerful means for examining the large sample form of a
statistical model (see, e.g., Abebe et al., \citeyear{Abeal95}; Andrews, Fraser and
Wong, \citeyear{AndFraWon05}; Cakmak et al., \citeyear{Caketal98}). From such expansions
we find that an asymptotic model to second order can be expressed as a
location model and to third order can be expressed as a location model
with an $O(n^{-1})$ adjustment that describes curvature.

Examples arise frequently in the vector parameter context. But for the
scalar parameter context the common familiar models are location or
scale models and thus without the curvature of interest here. A simple
example with curvature, however, is~the~gamma distribution model:
$f(y;\theta)=\break \Gamma^{-1}(\theta)y^{\theta-1}\exp\{-y\}$.

To illustrate the moderate curvature, we will take a~very simple
example where $y$ is $\operatorname{Normal} \{\theta, \sigma^2(\theta)\}$ and
$\sigma^2(\theta)$ depends just weakly on the mean $\theta$, and then
in asymptotic standardized form we would have
\[
\sigma^2(\theta) = 1 + \gamma\theta^2/2n
\]
in moderate deviations. The $\beta$-level quantile for this normal
variable $y$ is
%
\begin{eqnarray} \label{A}
y_\beta(\theta) & = & \theta+\sigma(\theta)z_\beta\nonumber\\
& = & \theta+ z_\beta(1+\gamma\theta^2/2n)^{1/2}\\
& = & \theta+ z_\beta(1+\gamma\theta^2/4n)+O(n^{-3/2}).\nonumber
\end{eqnarray}

The confidence bound $\hat\theta_\beta$ with $\beta$ confidence above
can be obtained from the usual Fisher inversion of $y=\theta+z_\beta(1+
\gamma\theta^2/4n)$: we obtain
\begin{eqnarray*}
\theta& = & y - z_\beta(1+\gamma\theta^2/4n)+ O(n^{-3/2}) \\
& = & y - z_\beta\{1+ \gamma(y-z_\beta)^2/4n\}+O(n^{-3/2}).
\end{eqnarray*}
Thus, the $\beta$ level lower confidence quantile to order
$O(n^{-3/2})$ is
%
\begin{equation} \label{B}
\hat\theta^C(y) = y - z_\beta\{1+ \gamma(y-z_\beta)^2/4n\},
\end{equation}
where we add the label $C$ for confidence to distinguish it from other
bounds soon to be calculated. See Figure~\ref{F}.

(ii) \textit{From confidence to likelihood}. We are interested in
examining posterior quantiles for the adjusted normal model and in this
section work from the confidence quantile to the likelihood quantile,
that is, to the posterior quantile with flat prior\break $\pi(\theta)=1$;
this route seems computationally easier than directly calculating a
likelihood integral.

From Section \ref{sec3} and formula (\ref{A}) above, we have that the prior
$\pi(\theta)$ that converts a likelihood $f^L(\theta)=\allowbreak L(\theta; y^0)\,{=}\,F_y(y^0; \theta)$
to confidence $f^C(\theta)\,{=}\,-F_{;\theta}(y^0;\theta)$ is
\begin{eqnarray*}
\frac{dy}{d\theta} \bigg|_{y^0} & = & 1+ \gamma z\theta/2n|_{y^0 }\\
& = & 1+ \gamma(y^0-\theta)\theta/2n+O(n^{-3/2}) \\
& = & \exp\{\gamma(y^0-\theta)\theta/2n\}+O(n^{-3/2}).
\end{eqnarray*}
Then to convert in the reverse direction, from confidence
$f^C(\theta)$ to likelihood $f^L(\theta)$, we need the inverse weight
function
%
\begin{equation}\label{C}
w(\theta)= \exp\{\gamma\theta(\theta-y^0)/2n\}.
\end{equation}
Interestingly, this function is equal to $1$ at~$\theta=0$ and at~$y^0$, and is less than $1$ between these points when $\gamma>0$.

%
\begin{figure*}

\includegraphics{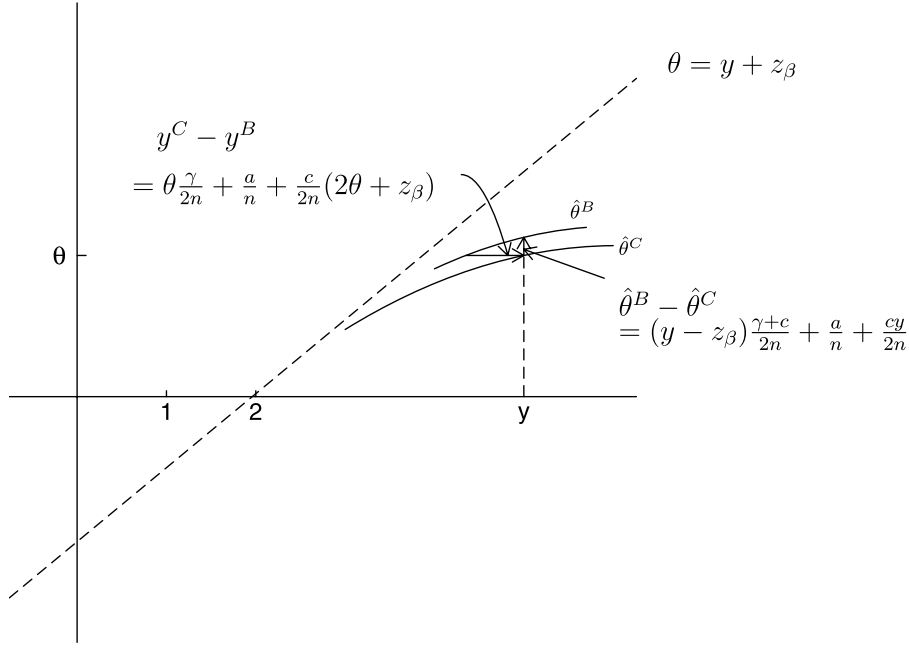}

\caption{$\beta$-level quantiles. The difference
$\hat\theta^B(y)-\hat\theta^C(y)$ is the vertical separation above $y$
between quantile curves. The difference $y^C(\theta)-y^B(\theta)$ is
the horizontal separation between curves as a function of $\theta$.}\label{G}
\vspace*{6pt}
\end{figure*}

(iii) \textit{From confidence quantile to likelihood quantile}. The weight
function (\ref{C}) that converts confidence to likelihood has the form
$\exp\{a\theta/n^{1/2}+c\theta^2/\allowbreak 2n\}$ with $a=-\gamma
y^0/2n^{1/2}$ and $c=\gamma$. The effect of such a tilt and bending is
recorded in the \hyperref[sec:Intrinsic continuity]{Appendix}. The\vspace*{1pt} confidence\vspace*{2pt} quantile $\hat\theta^C_\beta$
given at (\ref{B}) is a $1-\beta$ quantile of the confidence
distribution. Then using formula (\ref{A.2}) in the \hyperref[sec:Intrinsic continuity]{Appendix}, we obtain
the formula for converting confidence quantile to likelihood quantile:
%
\begin{eqnarray}\label{D}
\hat\theta^L & = & \hat\theta^C\biggl(1+ \frac{\gamma}{2n}\biggr) - \gamma
y^0 / 2n + \gamma y^0/2n \nonumber\\ [-8pt]\\ [-8pt]
& = & \hat\theta^C\biggl(1+\frac{\gamma}{2n}\biggr).\nonumber
\end{eqnarray}
Thus, the likelihood distribution is obtained from the confidence
distribution by a simple scale factor $1 + \gamma/2n$; this directly
records the consequence of the curvature added to the simple normal
model by having $\sigma^2(\theta)$ depend weakly on $\theta$.

(iv) \textit{From likelihood quantile to posterior quantile}. Now consider
a prior applied to the likelihood distribution. A prior can be expanded
in terms of standardized coordinates and takes the form
$\pi(\theta)=\exp(a\theta/n^{1/2} + c\theta^2/2n)$. The effect on
quantiles is available from the \hyperref[sec:Intrinsic continuity]{Appendix} and we see that a prior with
tilt coefficient $a/n^{1/2}$ would excessively displace the quantile
and thus would give posterior quantiles with bad behaving
$\operatorname{Propn}(\theta)$ in repetitions; accordingly, as a possible prior
adjustment, we consider a tilt with just a~coefficient $a/n$. We then
examine the prior $\pi(\theta)=\exp(a\theta/n + c\theta^2/2n)$.
First, we obtain the Bayes quantile in terms of the likelihood quantile
as
\[
\hat\theta^B=\hat\theta^L\biggl(1+\frac{c}{2n}\biggr) + \frac{a}{n} + \frac{cy}{2n};
\]
and then substituting for the likelihood quantile in terms of the
confidence quantile (\ref{D}) gives
%
\begin{equation}\label{bbb}
\hat\theta^B=\hat\theta^C\biggl(1+\frac{\gamma+ c}{2n}\biggr) + \frac{a}{n} +
\frac{cy}{2n}.
\end{equation}
For $\hat\theta^B(y)$ in (\ref{B}) to be equal to $\hat\theta^C(y)$ in
(\ref{bbb}) we
would need to have $c=-\gamma$ and then $a=\gamma y/2$. But this would
give a data dependent prior. We noted the need for data dependent
priors in Section \ref{sec3}, but we now have an explicit expression for the
effect of priors on quantiles.

Now consider the difference in quantiles:
\begin{eqnarray*}
\hat\theta^B(y)-\hat\theta^C(y) & = & \hat\theta^C
\biggl(\frac{\gamma+c}{2n} \biggr) + \frac{a}{n} + \frac{cy}{2n} \\
& = & (y-z_\beta) \frac{\gamma+c}{2n} + \frac{a}{n} + \frac{cy}{2n}\\
& = & \frac{a}{n} + y\frac{\gamma+ 2c}{2n} - z_\beta\frac{\gamma+c}{2n},
\end{eqnarray*}
where we have replaced $\hat\theta^C$ by $y-z_\beta$, to order\break
$O(n^{-3/2})$; Figure \ref{G} shows this difference as the vertical
separation above a data value $y$. From the third expression above we
see that in the presence of model curvature $\gamma$ the Bayesian
quantile can achieve the quality of confidence only if the prior is
data dependent or dependent on the level $\beta$.

Similarly, we can calculate the horizontal separation corresponding to
a $\theta$ value, and obtain
%
\begin{eqnarray}\label{new1}
\hspace*{7pt}y^C(\theta)-y^B(\theta) & =&\theta\frac{\gamma+ c}{2n} + \frac{a}{n} + \frac{c}{2n}(\theta+ z_\beta)
\nonumber\\ [-8pt]\\ [-8pt]
\hspace*{7pt}&=& \theta\frac{\gamma}{2n} + \frac{a}{n} + \frac{c}{2n}(2\theta+
z_\beta).\nonumber
\end{eqnarray}
This gives the quantile difference, the confidence quantile less the
Bayes quantile, as a function of $\theta$; see Figure \ref{G}, and
observe the horizontal separation to the right of a parameter value
$\theta$.

%
\begin{figure}

\includegraphics{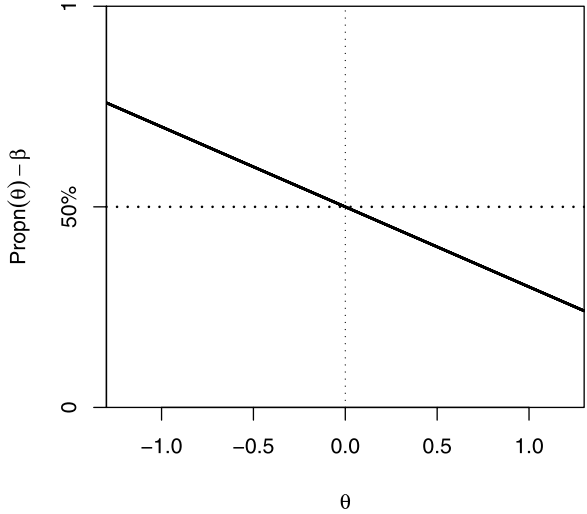}

\caption{The actual Proportion with claimed level $\beta=50\%$.}
\vspace*{10pt}
\label{Ex3_1}
\end{figure}

A Bayes quantile can not generate true statements concerning a
parameter with the reliability of confidence unless the model
curvature is zero, that is, unless the model is of the special location
form where Bayes coincides confidence. The Bayes approach can thus
be viewed as having a long history of misdirection.

\begin{figure}

\includegraphics{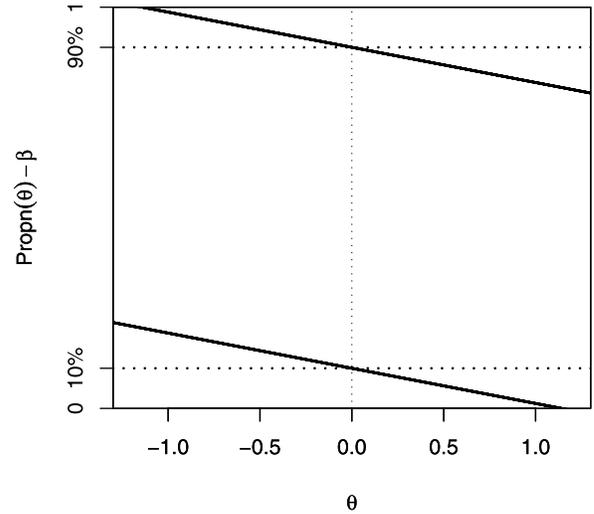}

\caption{The actual Proportion with claimed levels
$\beta= 90\%$ and $\beta=10\%$.} \label{Ex3_2}
\end{figure}

Now let $\theta$ designate the true value of the parameter $\theta$,
and suppose we examine the performance of the Bayesian and frequentist
posterior quantiles. In repetitions the actual proportion of instances
where $y < y^C(\theta)$ is of course $\beta$. The actual proportion of
cases with $y<y^B(\theta)$ is then
\[
\operatorname{Propn} (\theta)=\beta-\biggl\{\theta\frac{\gamma}{2n}
+ \frac{a}{n} + \frac{c}{2n}(2\theta+ z_\beta)\biggr\}\phi(z_\beta),
\]
where for the terms of order $O(n^{-1})$ it suffices to use the
$N(\theta, 1)$ distribution for $y$. The Bayes calculation claims the
level $\beta$. The choice $a=0, c=0$ gives a flat prior in the
neighborhood of $\theta=0$ which is the central point of the model
curvature. With such a choice the actual Proportion from the Bayes
approach is deficient by the amount $\theta\gamma\phi(z_\beta)/2n$. For
a claimed $\beta=50\%$ quantile see Figure \ref{Ex3_1} for the actual
Proportion and for a claimed $\beta=90\%$ or $\beta=10\%$ see Figure
\ref{Ex3_2}. Thus, the $\beta$ quantile by Bayes is consistently below
the claimed level $\beta$ for positive values of $\theta$, and
consistently above the claimed level for negative values of~$\theta$.

In summary, even in the scalar parameter context, an elementary
departure from simple linearity can lead to a performance error for the
Bayes calculation of asymptotic order $O(n^{-1})$. And, moreover, it is
impossible by the Bayes method to duplicate the standard confidence
bounds: a stunning revelation!

%

\section{The Paradigm}\label{sec7}

The Bayes proposal makes critical use of the conditional probability
formula $f(y_1|y^0_2)=cf(y_1,y_2^0)$. In typical applications the
formula has variables $y_1$\break and~$y_2$ in a temporal order: the value
of the first~$y_1$ is inaccessible and the value of the second $y_2$ is
observed with value, say, $y_2^0$. Of course, the value of the first~$y_1$
has been realized, say, $y_1^r$, but is concealed and is
unknown. Indeed, the view has been expressed that the only
probabilities possible concerning such an unknown $y_1^r$ are the
values $0$ or $1$ and we don't know how they would apply to that
$y_1^r$. We thus have the situation where there is an unknown
constant~$y_1^r$, a constant that arose antecedent in time to the
observed value~$y_2^0$, and we want to make probability statements
concerning that unknown antecedent constant. As part of the temporal
order we also have that the joint density became available in the
order~$f(y_1)$ for the first variable followed by $f(y_2|y_1)$ for the
second; thus, $f(y_1,y_2^0)= f(y_1) f(y^0_2|y_1)$.

The conditional probability formula itself is very much part of the
theory and practice of probability and statistics and is not in
question. Of course, limit operations are needed when the condition
$y_2=y_2^0$ has probability zero leading to a conditional probability
expression with a zero in the denominator, but this is largely
technical.

A salient concern seemingly centers on how probabilities can
reasonably be attached to a constant that is concealed from view? The
clear answer is~in terms of what \textit{might} have occurred given the
same observational information: the corresponding picture is of many
repetitions from the joint distribution giving pairs $(y_1,y_2)$;
followed by selection of pairs that have exact or approximate
agreement $y_2=y_2^0$; and then followed by examining the pattern in
the $y_1$ values among the selected pairs. The pattern records what
would have occurred for $y_1$ among cases where $y_2=y_2^0$; the
probabilities arise both from the density $f(y_1)$ and from the density
$f(y_2|y_1)$. Thus, the initial pattern $f(y_1)$ when restricted to
instances where $y_2=y_2^0$ becomes modified to the pattern $f(y_1|
y_2^0)=\allowbreak cf(y_1,y_2^0)= cf(y_1) f(y^0_2|y_1)$.

Bayes (\citeyear{Bay63}) promoted this conditional probability formula and its
interpretation, for statistical contexts that had no preceding
distribution for $\theta$ and he did so by introducing the
mathematical prior. He did provide, however, a motivating analogy and
the analogy did have something extra, an objective and real
distribution for the parameter, one with probabilities that were well
defined by translational invariance. Such a use of analogy in science
is normally viewed as wrong, but the needs for productive methodology
were high at that time.

If $\pi(\theta)$ is treated as being real and descriptive of how the
value of the parameter arose in the application, it would follow that
the preceding conditional probability analysis would give the
conditional description
\begin{eqnarray*}
\pi(\theta| y^0)&=&c\pi(\theta)f(y^0; \theta) \\
&=&c\pi(\theta)L^0(\theta).
\end{eqnarray*}
The interpretation for this would be as follows: In many repetitions
from $\pi(\theta)$, if each $\theta$ value was followed by a $y$ from
the model $f(y;\theta)$, and if the instances $(\theta, y)$ where $y$
is close to $y^0$ are selected, then the pattern for the corresponding
$\theta$ values would be $c\pi(\theta)L^0(\theta)$. In other words,
the initial relative frequency $\pi(\theta)$ for $\theta$ values is
modulated by $L^0(\theta)$ when we select using $y=y^0$; this gives the
modulated frequency pattern $c\pi(\theta)L^0(\theta)$. The conditional
probability formula as used in this context is often referred to as the
Bayes formula or Bayes theorem, but as a probability formula it long
predates Bayes and is generic; for the present extended usage it is
also referred to as the Bayes paradigm (Bernardo and Smith, \citeyear{BerSmi94}).

The Bayes' example as discussed in Sections \ref{sec2} and~\ref{sec3} examined a location
model $f(y-\theta)$ and the only prior that could represent location
invariance is the constant or flat prior in the location
parameterization, that is, $\pi(\theta)=c$. This of course does not
satisfy the probability axioms, as the total probability would be
$\infty$. The step, however, from just a~set of~$\theta$ values with
related model invariance to a~distribution for~$\theta$ has had the
large effect of emphasizing likelihood $L^0(\theta)$, as defined by
Fisher (\citeyear{FisN2}). And it has also had the effect, perhaps unwarranted,
of suggesting that the mathematical posterior distribution obtained
from the paradigm could be treated as a distribution of real
probability. If the parameter to variable relationship is linear, then
Section \ref{sec3} shows that the calculated values have the confidence (Fisher,
\citeyear{FisN2}; Neyman, \citeyear{Ney37}) interpretation. But if the relationship is
nonlinear, then the calculated numbers can seriously fail to have that
confidence property, as determined in Sections \ref{sec4}--\ref{sec6}; and indeed
fail to have anything with behavior resembling probability. The
mathematical priors, the invariant priors and other generalizations are
often referred to in the current Bayesian literature as objective
priors, a term that is strongly misleading.

In other contexts, however, there may be a real source for the
parameter $\theta$, sources with a known distribution, and thus fully
entitled to the term objective prior; of course, such examples do not
need the Bayes approach, they are immediately analyzable by probability
calculus. And, thus, to use objective to also refer to the
mathematical priors is confusing.

In short, the paradigm does not produce probabilities from no
probabilities. And if the required linearity for confidence is only
approximate, then the confidence interpretation can correspondingly be
just approximate. And in other cases even the confidence interpretation
can be substantially unavailable. Thus, to claim probability when even
confidence is not applicable does seem to be fully contrary to having
acceptable meaning in the language of the discipline.

\vspace*{-3pt}\section{Optimality}\label{sec8}\vspace*{-3pt}

Optimality is often cited as support for the Bayes approach: If we have
a criterion of interest that provides an assessment of a statistical
procedure, then optimality under the criterion is available using
a~procedure that is optimal under some prior average of the model. In
other words, if you want optimality,\vadjust{\goodbreak} it suffices to look for a
procedure that is optimal for the prior-average version of the model.
Thus, restrict one's attention to Bayes solutions and just find an
appropriate prior to work from. It sounds persuasive and it is
important.

Of course, a criterion as mentioned is just a numerical evaluation and
optimality under one such criterion may not give optimality under some
other criterion; so the choice of the criterion can be a major concern
for the approach. For example, would we want to use the length of a
posterior interval as the criterion or say the squared length of the
interval or some other evaluation; it makes a difference because the
optimality has to do with an average of values for the criterion and
this can change with change in the criterion.

The optimality approach can lead to interesting results but can also
lead to strange trade-offs; see, for example, Cox (\citeyear{Cox58}) and Fraser and
McDunnough (\citeyear{FraMcD80}). For if the model splits with known probabilities
into two or several components, then the optimality can create
trade-offs between these; for example, if data sometimes is high
precision and sometimes low precision and the probabilities for this
are available, then the search for an optimum mean-length confidence
interval at some chosen level can give longer intervals in the high
precision cases and shorter intervals in the low precision cases as a
trade-off toward optimality and toward intervals that are shorter on
average. It does sound strange but the substance of this phenomenon is
internal to almost all model-data contexts.

Even with a sensible criterion, however, and without the compound
modeling and trade-offs just mentioned, there are serious difficulties
for the optimality support for the Bayes approach. Consider further the
example in Section \ref{sec6} with a location $\operatorname{Normal}$ variable where the variance
depends weakly on the mean: $y$ is $\operatorname{Normal}\{\theta, \sigma^2(\theta
)\}$
with $\sigma^2(\theta)=1+\gamma\theta^2/2n$ and where we want a bound
$\hat\theta_\beta(y)$ for the parameter~$\theta$ with reliability
$\beta$ for the assertion that $\theta$ is larger than
$\hat\theta_\beta(y)$.

From confidence theory we have immediately (\ref{B}) that
\[
\hat\theta(y)=\hat\theta^C(y) = y - z_\beta\{1+
\gamma(y-z_\beta)^2/4n\}
\]
with $O(n^{-3/2})$ accuracy in moderate deviations.\break What is available
from the Bayes approach? A prior $\pi(\theta)=\exp
\{a\theta/n^{1/2} + c\theta^2/2n\}$ gives the posterior bound
\[
\hat\theta^\beta(y)=\hat\theta^C(y)\{1+c/2n\}+\frac{a}{n}+\frac{cy}{2n}.
\]
The actual Proportion for the $\beta$ level confidence bound is exactly
$\beta$. The actual Proportion, however, for the Bayes bound as
derived (\ref{new1}) is
\[
\beta-\biggl\{\theta{\gamma\over2n} + \frac{a}{n}+\frac{c}{2n}(2\theta+z_\beta)\biggr\}\phi(z_\beta);
\]
and there is no choice for the prior, no choice for $a$ and $c$, that
will make the actual equal to the nominal unless the model has nonzero
curvature $\gamma$.

We thus have that a choice of prior to weight the likelihood function
can not produce a $\beta$ level bound. But a $\beta$ level bound is
available immediately and routinely from confidence methods, which does
use more than just the observed likelihood function.

Of course, in the pure location case the Bayes approach is linear and
gives confidence. If there is nonlinearity, then the Bayes procedure
can be seriously inaccurate.

\section{Discussion}\label{sec9}

Bayes (\citeyear{Bay63}) introduced the observed likelihood function to general
statistical usage. He also introduced the confidence distribution when
the application was to the special case of a location model; the more
general development (Fisher, \citeyear{Fis30}) came much later and the present name
confidence was provided by Neyman (\citeyear{Ney37}). Lindley (\citeyear{Lin58}) then observed
that the Bayes derivation and the Fisher (\citeyear{Fis30}) derivation coincided
only for location models; this prompted continuing discord as to the
merits and validity of the two procedures in providing a~probability-type assessment of an unknown parameter value.

A distribution for a parameter value immediately 
makes available a quantile for that parameter, at any percentage level
of interest. This means that the merits of a procedure for evaluating
a parameter can be assessed by examining whether the quantile relates
to the parameter in anything like the asserted rate or level asserted
for that quantile. The examples in Sections \ref{sec4}--\ref{sec6} demonstrate that
departure from linearity in the relation between parameter and variable
can seriously affect the ability of likelihood alone to provide
reliable quantiles for the parameter of interest.

There is of course the question as to where the prior comes from and
what is its validity? The prior could be just a device as with Bayes
original proposal, to use the likelihood function directly to provide
inference statements concerning the parameter. This has been our
primary focus and such priors can reasonably be called default
priors.\vadjust{\goodbreak}

And then there is the other extreme where the prior describes the
statistical source of the experimental unit or more directly the
parameter value being considered. We have argued that these priors
should be called objective and then whether to use them to perform the
statistical
analysis is a reasonable question.

Between these two extremes are many variations such as subjective
priors that describe the personal views of an investigator and elicited
priors that represent some blend of the background views of those close
to a current investigation. Should such views be kept separate to be
examined in parallel with objective views coming directly from the
statistical investigation itself or should they be blended into the
computational procedure applied to the likelihood function alone? There
would seem to be strong arguments for keeping such information
separate from the analysis of the model with data; any user could then
combine the two as deemed appropriate in any subsequent usage of the
information.

Linearity of parameters and its role in the Bayesian frequentist
divergence is discussed in Fraser, Fraser and Fraser (\citeyear{Fra2010a}). Higher order
likelihood methods for Bayesian and frequentist inference were surveyed
in B{\'e}dard, Fraser and
Wong (\citeyear{BedFraWon07}), and an original intent there was to include
a comparison of the Bayesian and frequentist results. This, however,
was not feasible, as the example used there for illustration was of
the nice invariant type with the associated theoretical equality of
common
Bayesian and frequentist probabilities; thus, the anomalies discussed
in this paper were not overtly available there.

\section{Summary}\label{sec10}

A probability formula was used by Bayes (\citeyear{Bay63}) to combine a
mathematical prior with a model plus data; it gave just a mathematical
posterior, with no consequent objective properties. An analogy
provided by Bayes did have a real and descriptive prior, but it was
not part of the problem actually being examined.

A familiar Bayes example uses a~special model, a location
model; and the resulting intervals have attractive properties, as
viewed by many in statistics.

Fisher (\citeyear{FisN2}) and Neyman (\citeyear{Ney37}) defined confidence. And the Bayes
intervals in the location model case are seen to satisfy the confidence
derivation, thus providing an explanation for the attractive
properties.

The only source of variation available to support a~Bayes posterior
probability calculation is that provided by the model, which is what
confidence uses.

Lindley (\citeyear{Lin58}) examined the probability formula argument and the
confidence argument and found that they generated the same result only
in the Bayes location model case; he then judged the confidence
argument to be wrong.

If the model, however, is not location and, thus, the variable is not
linear with respect to the parameter, then a Bayes interval can
produce correct answers at a rate quite different from that claimed by
the Bayes probability calculation; thus, the Bayes posterior may be an
unreliable presentation, an unreliable approximation to confidence, and
can thus be judged as wrong.

The failure to make true assertions with a promised reliability can be
extreme with the Bayes use of mathematical priors (Stainforth et al.,
\citeyear{Staetal}; Heinrich, \citeyear{autokey20}).

The claim of a probability status for a statement that can fail to be
approximate confidence is misrepresentation. In other areas of science
such false claims would be treated seriously.

Using weighted likelihood, however, can be a fruitful way to explore
the information available from just a~likelihood function. But the
failure to have even a~confidence interpretation deserves more than
just gentle caution.

A personal or a subjective or an elicited prior may record useful
background to be recorded in parallel with a confidence assessment. But
to use them to do the analysis and just get approximate or biased
confidence seems to overextend the excitement of exploratory
procedures.

\begin{appendix}
\section*{Appendix}
\label{sec:Intrinsic continuity}

\renewcommand{\theequation}{\arabic{equation}}
\setcounter{equation}{8}
\subsection*{Tilting, Bending and Quantiles}

Consider a variable $y$ that has a $\operatorname{Normal} (\theta; 1)$ distribution
and suppose that its density is subject to an exponential tilt and
bending as described by the modulating factor $\exp\{ay +
cy^2/2\}$. It follows easily by completing the square in the exponent
that the new variable, say, $\tilde{y}$, is also normal but with mean
$(\theta+a)/(1-c)$ and variance $1/(1-c)$. In particular, we can write
\[
\tilde{y}=\frac{\theta+ a}{1-c} + (1-c)^{-1/2}z,
\]
where $z$ is standard normal. And if we let $z_\beta$ be the~$\beta$\vadjust{\goodbreak}
quantile of the standard normal with $\beta=\Phi(z_\beta)$, then the~$\beta$ quantile of $\tilde y$ is
\[
\tilde{y}_\beta= \frac{\theta+a}{1-c} + (1-c)^{-1/2}z_\beta.
\]
Thus, with the $\operatorname{Normal}(\theta, 1)$ we have that tilting and bending
just produce a location scale adjustment to the initial variable.

Now suppose that $y=\theta+z$ is $\operatorname{Normal} (\theta; 1)$ to third order,
and suppose further that its density receives an exponential tilting
and bending described by the factor $\exp \{ay/n^{1/2} +
cy^2/2n\}$. Then from the preceding we have that the new variable can
be expressed in terms of preceding variables as
%
\begin{eqnarray} \label{A.1}
\tilde{y} & = & {\theta+a/n^{1/2} \over1-c/n}+(1-c/n)^{-1/2}z\nonumber\\
& = & \theta(1+c/n)+a/n^{1/2} + (1+c/2n)z \\
& = & y(1+c/2n)+a/n^{1/2} + \theta c/2n,\nonumber
\end{eqnarray}
where succeeding lines use adjustments that are\break $O(n^{-3/2})$. The
second line on the right gives quantiles in terms of the standard
normal and the third line gives quantiles in terms of the initial
variable $y$.

One application for this arises with posterior distributions. Suppose
that $\theta=y^0 +z$ is $\operatorname{Normal} (y^0,1)$ to third order and that its
density receives a tilt and bending described by $\exp(a\theta/n^{1/2}
+ c\theta^2/2n)$. We then have from (\ref{A.1}) that the modified
variable can be expressed as
%
\begin{eqnarray}\label{A.2}
\hspace*{35pt}\tilde{\theta}
& = & y^0(1+c/n)+a/n^{1/2} + (1+c/2n)z \nonumber\\ [-8pt]\\ [-8pt]
\hspace*{35pt}& = & \theta(1+c/2n)+a/n^{1/2} + y^0 c/2n,\nonumber
\end{eqnarray}
to order $O(n^{-3/2})$.
\end{appendix}

\section*{Acknowledgments}

The author expresses deep appreciation to Nancy Reid for many helpful
discussions from various viewpoints; and we join to express deep appreciation to Pekka
Sinervo who introduced us to statistical concerns in High Energy Physics and then
took
valuable time from Decanal duties to clarify contextual issues. Also very special thanks go to Ye Sun,
Tara Cai and Kexin Ji for many contributions toward this research,
including numerical computation and preparation of figures and work on
the manuscript. The Natural Sciences and Engineering Research Council
of Canada has provided financial support. The author also thanks the
reviewers for very helpful comments.

%

\end{document}